\documentclass[useAMS,usenatbib,twocolumn]{mn2e}
\usepackage{amssymb}
\usepackage{amsmath}
\usepackage{graphicx}
\begin{document}

\title{Behavior of Torsional Alfven Waves and Field Line Resonance on
Rotating Magnetars}
\author[T. Okita and Y. Kojima]
{Taishi Okita \& Yasufumi Kojima \\
{\small Depertment of Physics, Hiroshima University, Higashi-Hiroshima
739-8526, Japan;}\\
{\small okita@hirax7.hepl.hiroshima-u.ac.jp,
kojima@theo.phys.sci.hiroshima-u.ac.jp}}
\date{}
\maketitle

\begin{abstract}
Torsional Alfven waves are likely excited with bursts in rotating magnetars.
These waves are probably propagated through corotating atmospheres toward a
vacuum exterior. We have studied the physical effects of the azimuthal wave
number and the characteristic height of the plasma medium on wave
transmission. In this work, explicit calculations were carried out based on
the three-layered cylindrical model. We found that the coupling strength
between the internal shear and the external Alfven modes is drastically
enhanced, when resonance occurs in the corotating plasma cavity. The spatial
structure of the electromagnetic fields in the resonance cavity is also
investigated when Alfven waves exhibit resonance.
\end{abstract}

\section{ Introduction}

Soft gamma-ray repeaters (SGRs) and anomalous X-ray pulsars (AXPs) have been
known as strongly magnetized neutron stars, \textit{magnetars}. The hallmark
of these objects is to repeat X-ray or gamma-ray emissions irregularly. So
far, four or five known active objects which show frequent X-ray emission of
tremendously energetic and shorter initial bursts (typically $E\lesssim
10^{41}$ ergs and $\Delta t\sim 0.5$ s) have been identified with SGRs
(Thompson \textit{et al.} 2001). Some of them occasionally exhibit more
energetic events. The giant flare 1900+14, now associated with SGR 0525-66,
was observed in 1979 (Mazets, Golenetskii, and Gur'yan 1979) for the first
time and became active again in 1992 (Kouveliotou \textit{et al.} 1993) and
also in 1998 (Kouveliotou \textit{et al.} 1998 ; Hurley \textit{et al}.
1999). Recently, the intense $\gamma $-ray flare from SGR 1806-20 occured on
December 27, 2004 was reported (Palmer \textit{et al}. 2005; Hurley \textit{%
et al}. 2005; Mereghetti \textit{et al}. 2005). Rea \textit{et al. }(2005)
have investigated the pulse profile and flare spectrum of SGR 1806-20. Their
study may potentially give the information about the global field structure
in the magnetosphere and may further promote theoretical magnetar models.
Their extream peak luminosity extends up to $10^{6}L_{\text{Edd}}$ estimated
at a distance of 10 kpc (Mazets \textit{et al}. 1999; Feroci \textit{et al}.
2001). Typical X-ray luminosities of SGRs have been measured to be $L_{\text{%
x}}=10^{34}$-$10^{36}$ erg s$^{-1}$ (Hurley \textit{et al.} 2000; Thompson 
\textit{et al.} 2000), except with giant bursts SGRs 1900+14 and 1806-20.
Shorter durations of initial flares of SGRs are comparable to the Alfven
crossing time of the core. The energy distribution shows good agreement with
Gutenberg-Lichter law, which may indicate statistical similarity to
earthquakes or solar flares (Cheng \textit{et al}. 1996; Gogus \textit{et al}%
. 1999, 2000). The SGRs have spin periods in a small range of $P=5$-$8$ s
with rapid spin down rate $\dot{P}\simeq 10^{-10}$ ss$^{-1}$ , which
therefore give characteristic ages $P/\dot{P}\sim $ $10^{3}$ yr (Mazets 
\textit{et al}. 1979; Kouveliotou \textit{et al}. 1998; Hurley \textit{et al}%
. 1999a).

On the other hand, the physical nature of AXPs is still uncertain due to
poor observations, but there seem to be some similarities and differences
between SGRs and AXPs. AXPs are energetic sources of pulsed X-ray emission,
whose periods lie in a narrow range $P=6$-$12$ s, characteristic ages $P/%
\dot{P}=$ $3\times 10^{3}\sim 4\times 10^{5}$ yr and X-ray luminosities $L_{%
\text{x}}=5\times 10^{34}$-$10^{36}$ erg s$^{-1}$(Mereghetti 2000; Thompson
2001). AXP sources likely have somewhat larger active ages and some of them
have softer X-ray spectra compared with SGRs. One of the primary difference
between them will be that AXPs so far have shown only quiescent X-ray
emission with no bright active bursts such as giant flares. Active ages of
SGRs and AXPs are consistent with the observed evidence that these compact
objects often give their location close to the edge of shell-type supernova
remnants. Neither SGRs nor AXPs show the presence of conspicuous
counterparts at other wavelengths (Mereghetti \& Stella 1995; Mereghetti 
\textit{et al}. 2002). Origin of these enourmous energetics is likely to
come from their strong magnetic fields $B=10^{14}$-$10^{15}$ G estimated by
their spin periods and spin down rates. Only the magnetar model has been
able to account for the enigmatic properties of a rare class of SGRs or AXPs.

Shear and Alfvenic waves play an essential role on the energy transfer to
the exterior in the burst-like phenomena observed in SGRs or giant flares.
The Alfven wave propagation in such a strongly magnetized star should be
clarified theoretically. In our previous work, the propagation and
transmission of torsional Alfven waves have been studied, focused only on
the fundamental mode of azimuthal wave number $m=1$ as a first step. In that
work, the exterior of the star is examined qualitatively only, assuming two
extreme cases: (i) corotating together with the star and (ii) static state
independent of the stellar rotation. The former will come true when the
plasma gas is trapped by closed magnetic field lines, and the latter will be
realized without any other force. In any case, it might be inadequate at
least in the following two points to place such strong constraints on the
wave mode and on ambient circumstances of the star in the previous paper.
First, under the more realistic situations various modes with azimuthal wave
number $m\gg 1$ are probably triggered by star quake. Second, it is natural
to assume that a scale height of the plasma gas is neither $L=0$ nor $%
L\rightarrow \infty $, but has a finite value of $L$. Huang \textit{et al}.
(1998) pointed out that a relativistic \textit{fireball} like those in
classical GRBs may exist in SGRs. Recently, Thompson and Duncan (2001) have
also suggested that after the initial hard spike emission, some lumps of hot
plasma gas involving electron-positron pairs and high energy photons, that
is, \textit{fireball, }would be created on closed magnetic field lines. In
fact, the fast decline and complete evaporation on X-ray light curve
observed in the August 27 burst provides a clear evidence of the trapped 
\textit{fireball}. Motivated by this, we thus extend our study to include
some plasma gas spreading over the stellar surface.

The main aim of this paper is to study the physical behavior of the
torsional Alfven waves with azimuthal wave number $m$ in the presence of
corotating plasma confined within a certain finite distance $L$ and then to
investigate the effects of these quantities $m$ and $L$ on the wave
propagation and transmission. In section 2, our model is self-consistently
constructed and the relevant basic equations are formulated. These equations
are almost the same as those derived in Kojima and Okita (2004), but are
summarized here for the paper to be self-contained. In section 3, we shall
derive the dispersion relation in a WKB way to show that the rotating plasma
atmosphere plays a crucial role as a resonant cavity of the wave when a
certain condition holds. In section 4, transmission rates of the Alfven
waves are numerically calculated. In section 5, their electromagnetic field
structure is also discussed with the numerical results. In section 6, we
give a brief summary of our findings and their implications for the
torsional Alfven waves on rotating magnetars.\bigskip\ 

\section{Electromagnetics on Rotating Magnetars}

\subsection{Model}

Both the magnetic field and the rotation of star lead to a complicated
geometrical configuration. In this paper we assume some simplified
conditions to understand the physical processes of the wave propagation. We
here consider three-layered cylindrical model with radius $\varpi _{\text{pc}%
}$. It is composed of the neutron star crust between $z=-q$ and $z=0$
(region (1)), corotating plasma above the stellar surface between $z=0$ and $%
z=L$ (region (2)) and static pure vacuum at $z>L$ (region (3)) as shown in
Fig.1. The plasma gas filled in the atmosphere corotates with the crust at
the same angular velocity $\Omega $. Local magnetic fields permeate
uniformly in each layer and point along the $z$ direction, $\mathbf{B}=B_{%
\text{o}}\mathbf{e}_{z}$, which represents open magnetic fields extending to
infinity.

Alfven waves excited at the bottom of the crust $q\approx 10^{5}$ cm owing
to some mechanisms, travel upward along the local magnetic field lines. They
are partially reflected and transmitted at the boundaries $z=0$ and $z=L$,
where the physical property of Alfven waves significantly changes because of
the effect of the background rotation, as shown below. This model, by
setting $L\rightarrow \infty $, reduces to the case in which the plasma gas
extends infinitely to the exterior of the star, while setting $L\rightarrow
0 $ reduces to the simple case in which the exterior is filled with pure
vacuum only. Up to the present time, we have no observational information of
the plasma size $L$. In this work, $L$ is regarded as a free parameter in
order to investigate its effect on the wave propagation and
transmission.
\begin{figure}
\includegraphics[scale=1.0]{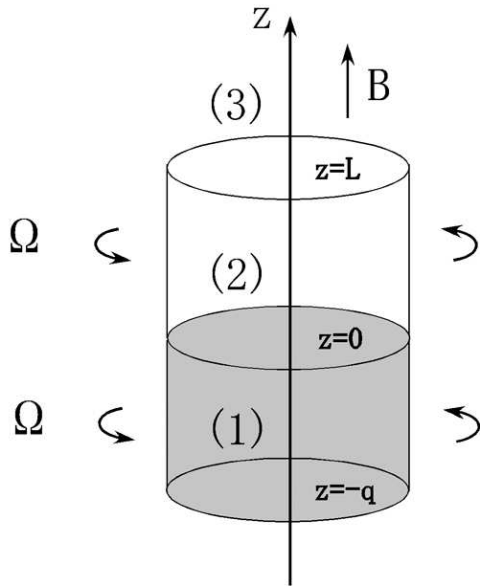}
\caption{
Three-layered
cylindrical model. Region (1) corresponds to the rotating neutron star crust
with a constant angular velocity 
$\Omega $, region (2)
corresponds to the \textit{fireball} as a resonant cavity filled with plasma
corotating with $\Omega $, and region (3) denotes a static
pure vacuum. 
}
\end{figure}
We shall now give some comments on the validity of this model by comparing
the physical sizes $\varpi _{\text{pc}}$, $q$ and $L$ with the stellar
radius $R$. In this model we have explicitly assumed that the local magnetic
field has a $z$-component only. Our model can be applied to the polar cap
region, whose cylindrical radius $\varpi _{\text{pc}}$ is given by $\varpi _{%
\text{pc}}=R\sin \theta _{\text{pc}}\approx 10^{4}\left( T/1\text{s}\right)
^{-1/2}$ cm $\ll $ $R=10^{6}$ cm. Curvature of the stellar surface and the
magnetic field lines may be neglected within the polar cap region. In a
similar way, it may be valid to assume that the local magnetic field lines
are uniform if the thickness of each layer is smaller than the star radius, $%
L,q\ll R$. We can also extend our model to the extreme case $L\sim R$. Even
in this case, the validity of this model nevertheless holds good near the
z-axis. In the remainder of this paper we ristrict our explicit calculations
only to the axially symmetric small region within the polar cap radius,
unless otherwise stated.

\subsection{Linear Perturbation}

In this section, we consider the propagation of torsional shear-Alfven waves
with various azimuthal modes on rotating magnetars. Such waves are probably
excited by the turbulent motion of the starquake in the deep crust, but
above the neutron drip $(z>z_{\text{nd}}\approx -10^{5}$ cm$)$. Many
proposals for starquake model have been put forward (e.g., Pacini and
Ruderman 1974), but all of them are generally argued only for weakly
magnetized neutron stars with $10^{11}-10^{12}$ G. Therefore, their
treatments are inadequate for magnetars, as they are. However, analogous
mechanism may also occur in magnetars. In this paper, we do not discuss the
triggering mechanism of the starquake itself, but focus only on the process
by which electromagnetic shear waves, once generated, are propagated and
transmitted from the deep crust, through a magnetized plasma, toward the
vacuum exterior.

We assume that the horizontal Lagrange displacement $\mathbf{\xi }=(\xi
_{\varpi },\xi _{\phi },0)$ is suddenly shaken in the deep interior at depth 
$q\approx 10^{5}$ cm, nevertheless the matter remains immobile in the
vertical direction due to strong gravity of the neutron star, $g\sim 10^{14}$
cm s$^{-2}$. Therefore the following transverse wave condition may be easily
satisfied.%
\begin{equation}
\nabla \cdot \mathbf{\xi =}\dfrac{1}{\varpi }\dfrac{\partial }{\partial
\varpi }\left( \varpi \xi _{\varpi }\right) +\dfrac{1}{\varpi }\dfrac{%
\partial \xi _{\phi }}{\partial \phi }=0.
\end{equation}%
As one of the simplest solutions, wave ansatz under this condition can be
written formally as%
\begin{align}
\mathbf{\xi }(\varpi ,\phi ,z,t)& =\left( \mathbf{e}_{\varpi }\pm i\mathbf{e}%
_{\phi }\right) \left( \dfrac{\varpi }{\varpi _{\text{pc}}}\right) ^{m-1}\xi
_{\pm m,\omega }\left( z\right) e^{-i\left( \omega t\mp m\phi \right) }\text{
}  \notag \\
(m& =1,2,3\cdots ).  \label{displacement}
\end{align}%
Here notation `$\pm $' denotes the difference of helicity states.
Displacements and other perturbed quantities are always assumed to have a
harmonic time dependency $e^{-i\omega t}$.

The crust and the surrounding plasma can be regarded as a perfect conductor
because ohmic dissipation timescales are very long compared with other
timescales of interest. The frozen condition can read%
\begin{equation}
\mathbf{E+}\dfrac{1}{c}\mathbf{v}\times \mathbf{B=}0.  \label{ff}
\end{equation}%
Background is assumed to rotate uniformly with $\mathbf{v}=\varpi \Omega 
\mathbf{e}_{\phi }$, which means that the local electric field arround a
highly conducting spherical star has only a radial component $\mathbf{E}%
=(-\varpi \Omega B_{\text{o}}/c)\mathbf{e}_{\varpi }$. The electric field
now generates the Goldreich-Julian charge density%
\begin{equation}
\rho _{\text{e}}=\nabla \cdot \mathbf{E}/4\pi =-\Omega B_{\text{o}}/2\pi c.
\label{Goldreich}
\end{equation}

Transverse displacements of the crustal matter produce electromagnetic field
perturbations propagating along the local magnetic field $\mathbf{B}=B_{%
\text{o}}\mathbf{e}_{z}$. A self-consistent set of Maxwell's equations for
perturbed electromagnetic fields $\delta \mathbf{E}$ and $\delta \mathbf{B}$
are 
\begin{gather}
\nabla \cdot \delta \mathbf{E}=4\pi \delta \rho _{\text{e}},  \label{max1} \\
\nabla \cdot \delta \mathbf{B}=0,  \label{max2} \\
\nabla \times \delta \mathbf{E}=-\dfrac{1}{c}\partial _{t}\delta \mathbf{B},
\label{max3} \\
\nabla \times \delta \mathbf{B}=\dfrac{4\pi }{c}\delta \mathbf{j}+\dfrac{1}{c%
}\partial _{t}\delta \mathbf{E,}  \label{max4}
\end{gather}%
where $\delta $ denotes Eulerian perturbation. By taking the perturbation of
equation (\ref{ff}) and then combining with equation (\ref{max3}), $\delta 
\mathbf{E}$ and $\delta \mathbf{B}$ are expressed as%
\begin{align}
\delta \mathbf{E}& =-\dfrac{1}{c}\left( \delta \mathbf{v}\times \mathbf{B}+%
\mathbf{v}\times \delta \mathbf{B}\right) ,  \label{dE1} \\
\delta \mathbf{B}& =\left( \mathbf{B}\cdot \nabla \right) \mathbf{\xi .}
\label{dB1}
\end{align}%
By using the relation between the displacement and the velocity perturbation 
$\delta \mathbf{v}=\partial _{t}\mathbf{\xi }+\left( \mathbf{v}\cdot \nabla
\right) \mathbf{\xi }-\left( \mathbf{\xi }\cdot \nabla \right) \mathbf{v}$
together with equation (\ref{displacement}), one obtains

\begin{align}
\delta \mathbf{E}& =\dfrac{B_{\text{o}}}{c}\left( \dfrac{\varpi }{\varpi _{%
\text{pc}}}\right) ^{m-1}  \notag \\
& \times \left[ \mp \left( \omega \mp m\Omega \right) \xi _{\pm }\left( 
\mathbf{e}_{\varpi }\pm i\mathbf{e}_{\phi }\right) +\Omega \varpi \dfrac{%
d\xi _{\pm }}{dz}\mathbf{e}_{z}\right] e^{-i\left( \omega t\mp m\phi \right)
},  \label{dE2} \\
\delta \mathbf{B}& =B_{\text{o}}\left( \dfrac{\varpi }{\varpi _{\text{pc}}}%
\right) ^{m-1}\dfrac{d\xi _{\pm }}{dz}\left( \mathbf{e}_{\varpi }\pm i%
\mathbf{e}_{\phi }\right) e^{-i\left( \omega t\mp m\phi \right) }.
\label{dB2}
\end{align}%
Equations (\ref{dE2}) and (\ref{dB2}) imply that if stellar rotation can be
completely ignored, then $\delta \mathbf{E}$, $\delta \mathbf{B}$ and the
propagation vector $\mathbf{k}=k\mathbf{e}_{z}$ form a mutually orthogonal
set of vectors, say, TEM modes. However, in the presence of rotation, such
an orthogonality breaks down and thus longitudinal modes of the perturbed
electric fields and currents are excited (TM modes). Other perturbed
quantities related to the displacement are calculated by using above the
results (\ref{dE2}) and (\ref{dB2}) 
\begin{align}
\delta \mathbf{j}& =\dfrac{c}{4\pi }\nabla \times \delta \mathbf{B}-\dfrac{1%
}{4\pi }\partial _{t}\delta \mathbf{E}  \notag \\
& =\mp i\dfrac{cB_{\text{o}}}{4\pi }\left( \dfrac{\varpi }{\varpi _{\text{pc}%
}}\right) ^{m-1}\times  \notag \\
& \left[ \left\{ \dfrac{d^{2}\xi _{\pm }}{dz^{2}}+\dfrac{\omega \left(
\omega \mp m\Omega \right) }{c^{2}}\xi _{\pm }\right\} \left( \mathbf{e}%
_{\varpi }\pm i\mathbf{e}_{\phi }\right) \mp \dfrac{\omega \Omega \varpi }{%
c^{2}}\dfrac{d\xi _{\pm }}{dz}\mathbf{e}_{z}\right]  \notag \\
& \times e^{-i\left( \omega t\mp m\phi \right) }, \\
\delta \rho _{\text{e}}& =\dfrac{1}{4\pi }\nabla \cdot \delta \mathbf{E} 
\notag \\
& =\dfrac{B_{\text{o}}\Omega \omega }{4\pi c}\left( \dfrac{\varpi }{\varpi _{%
\text{pc}}}\right) ^{m-1}\dfrac{d^{2}\xi _{\pm }}{dz^{2}}e^{-i\left( \omega
t\mp m\phi \right) }.
\end{align}

The linearized equation of motion for uniformly rotating background from a
static equilibrium state governing the torsional Alfven waves is now given by%
\begin{eqnarray}
&&\rho \left[ \partial _{t}\delta v_{i}+\left( \mathbf{v}\cdot \nabla
\right) \delta v_{i}+\left( \delta \mathbf{v}\cdot \nabla \right) v_{i}%
\right]  \notag \\
&=&\delta S_{i}+\delta F_{i}+\left( \mathbf{g}\delta \rho \right)
_{i}-\left( \nabla \delta p\right) _{i}\quad (i=\varpi ,\phi ,z).
\label{motion}
\end{eqnarray}%
Profile of the mass density $\rho $ in the crust is obtained by solving the
equation of state for degenerate electrons as follows (Blaes \textit{et al. }%
1989)%
\begin{align}
\rho & =\dfrac{\left( \mu _{\text{e}}m_{\text{u}}\right) ^{5/2}}{3\pi
^{2}\hbar ^{3}}\left( \dfrac{g^{2}\mu _{\text{e}}m_{\text{u}}}{c^{2}}%
z^{2}+2gm_{\text{e}}\left| z\right| \right) ^{3/2}  \notag \\
& \approx 8.0\times 10\left[ \left( \dfrac{\left| z\right| }{\text{cm}}%
\right) +2.5\times 10^{-4}\left( \dfrac{\left| z\right| }{\text{cm}}\right)
^{2}\right] ^{3/2}\text{g cm}^{-3}\text{,}  \label{mass}
\end{align}%
where $m_{\text{u}}$ is an atomic mass unit, $\mu _{\text{e}}$ is the mean
molecular weight per electron and others have usual physical meanings.

The first term of RHS in equation (\ref{motion}), $\delta S_{i}$, is a
perturbed elastic stress tensor associated with distorted matter in the
crust and can be written in terms of $\xi $%
\begin{align}
\delta S_{i}& =\nabla _{j}\left[ \left( \varkappa -\dfrac{2\mu }{3}\right)
\delta _{ij}\nabla \cdot \mathbf{\xi }\right] +\nabla _{j}\left[ \mu \left( 
\dfrac{\partial \xi ^{i}}{\partial x_{j}}+\dfrac{\partial \xi ^{j}}{\partial
x_{i}}\right) \right]  \notag \\
\quad (i,j& =\varpi ,\phi ,z),  \label{stress}
\end{align}%
where $\varkappa $ is a bulk modulus, $\mu $ is a shear modulus given by%
\begin{align}
\mu & =0.295Z^{2}e^{2}n_{i}^{4/3}  \notag \\
& \approx 4.8\times 10^{27}\left( \dfrac{\rho }{10^{11}\text{ g cm}^{-3}}%
\right) ^{4/3}\text{ erg cm}^{-3},  \label{shear}
\end{align}%
with the ion number density $n_{i}=\rho /Z\mu _{\text{e}}m_{\text{u}}$ and $%
\delta _{ij}$ Kronecker's delta. In the following calculations, $Z=32$ will
be adopted as a typical value in the crust. If one considers the complete
transverse\ oscillation mode, the first term of equation (\ref{stress})
vanishes owing to $\nabla \cdot \mathbf{\xi }=0$.

The second term on RHS of equation (\ref{motion}) can be resolved into
perturbations of Lorentz and Coulomb forces%
\begin{align}
\delta \mathbf{F}& =\delta \rho _{\text{e}}\mathbf{E+}\rho _{\text{e}}\delta 
\mathbf{E+}\dfrac{1}{c}\left( \delta \mathbf{j}\times \mathbf{B+j}\times
\delta \mathbf{B}\right)  \notag \\
& \simeq \dfrac{1}{c}\left( \delta \mathbf{j-}\rho _{\text{e}}\delta \mathbf{%
v}\right) \times \mathbf{B.}
\end{align}%
Here we used equation (\ref{dE1}) and unperturbed current $\mathbf{j=}\rho _{%
\text{e}}\mathbf{v}$ induced by the background rotation. We further dropped
the term $\delta \rho _{\text{e}}\mathbf{E}\propto \left( \Omega \varpi
/c\right) ^{2}$, which is small near the z-axis or within the actual stellar
radius $R$. Eliminating $\delta \mathbf{j}$ with the help of equation (\ref%
{max4}), the electromagnetic force reduces to 
\begin{equation}
\delta \mathbf{F}\simeq \dfrac{1}{4\pi }\left( \mathbf{B}\cdot \nabla
\right) ^{2}\mathbf{\xi }-\dfrac{B_{\text{o}}^{2}}{4\pi c^{2}}\partial
_{t}\delta \mathbf{v}.  \label{em force}
\end{equation}%
The first term in equation (\ref{em force}) implies the tension of the
perturbed magnetic field, while the second one shows the magnetic pressure
generated by the distorted matter.

The last two terms of equation (\ref{motion}) represent gravitational force $%
\left( \mathbf{g}\delta \rho \right) _{i}$ and pressure gradient $\left(
\nabla \delta p\right) _{i}$, respectively. However, unless one takes the p-
or f-mode such as compressional waves and/or sonic waves into consideration,
one can ignore them for simplicity in this model.

\subsection{Wave Equation}

We shall now derive the wave equation in region (1) shown in Fig.1.
Substituting equations (\ref{stress}) and ($\ref{em force}$) into equation
of motion (\ref{motion}), one can obtain the following differential wave
equation in terms of the displacement $\xi _{\pm }^{(1)}$ 
\begin{equation}
\dfrac{d^{2}\xi _{\pm }^{(1)}}{dz^{2}}+\dfrac{1}{\widetilde{\mu }}\dfrac{d%
\widetilde{\mu }}{dz}\dfrac{d\xi _{\pm }^{(1)}}{dz}+\dfrac{\widetilde{\rho }%
}{\widetilde{\mu }}\sigma _{\pm }\left[ \sigma _{\pm }\pm \left\{
m(1-h)+2\right\} \Omega \right] \xi _{\pm }^{(1)}=0,  \label{wave eq 1}
\end{equation}%
where $\widetilde{\mu }$ and $\widetilde{\rho }$ denote the effective shear
modulus and the effective mass density defined as%
\begin{gather}
\widetilde{\mu }=\mu +\dfrac{B_{o}^{2}}{4\pi }, \\
\widetilde{\rho }=\rho +\dfrac{B_{o}^{2}}{4\pi c^{2}}.
\end{gather}%
The ratio of these quantities gives the shear-Alfven wave speed in the crust 
$\widetilde{v}=\sqrt{\widetilde{\mu }/\widetilde{\rho }}.$ In the above
expression, the frequency $\sigma _{\pm }\equiv \omega \mp m\Omega $
measured in corotating frame for each helicity state has been introduced. We
can limit this frequency to the positive regime $\sigma _{\pm }>0$ for the
symmetry $\xi _{\pm m,-\omega }=\xi _{\mp m,\omega }^{\ast }$. Dimensionless
function $h$ in equation (\ref{wave eq 1}) is formally defined as 
\begin{equation}
h\equiv \dfrac{4\pi \rho c^{2}}{4\pi \rho c^{2}+B_{\text{o}}^{2}},
\end{equation}%
which has a great influence on the dispersion relation of the wave
especially with large $m$ not only in the inner surface, but also in the
rotating plasma cloud, as discussed in the following section. Note that if
one considers the static background $(\Omega =0)$, wave equation (\ref{wave
eq 1}) coincides with the one already derived by Blaes \textit{et al}.
(1989). We now look for a WKB solution. In the deep interior the solution of
wave equation (\ref{wave eq 1}) can be well asymptotically given by%
\begin{eqnarray}
\xi _{\pm }^{\text{asymp}} &\approx &\left| z\right| ^{\beta }\left\{ A_{\pm
}\exp \left[ -i(\psi _{\pm }(z)+\omega t)\right] \right.  \notag \\
&&\left. +B_{\pm }\exp \left[ i(\psi _{\pm }(z)-\omega t)\right] \right\} ,
\label{asy solution}
\end{eqnarray}%
with $\beta =-7/4.$ Here $\psi _{\pm }(z)$ denotes eikonals defined as%
\begin{equation}
\psi _{\pm }(z)\equiv \int_{-q}^{z}dz^{^{\prime }}\dfrac{\sqrt{\sigma _{\pm }%
\left[ \sigma _{\pm }\pm \left\{ m\left( 1-h\right) +2\right\} \Omega \right]
}}{\widetilde{v}},  \label{eikonal}
\end{equation}%
for each mode. The first and second terms in equation (\ref{asy solution})
represent the upward-propagating Alfven wave with a complex incident
amplitude $A_{\pm }$ and a downward-propagating wave with a complex
reflection amplitude $B_{\pm }$ bounced at the stellar surface, respectively.

We now turn to the wave behavior in region (2) $(0<z<L).$ Mass density in
this region is so small that one can formally take the limit $h\rightarrow 0$%
. Thus equation (\ref{wave eq 1}) reduces to 
\begin{equation}
\dfrac{d^{2}\xi _{\pm }^{(2)}}{dz^{2}}+\dfrac{1}{c^{2}}\sigma _{\pm }\left[
\sigma _{\pm }\pm (m+2)\Omega \right] \xi _{\pm }^{(2)}=0.  \label{wave eq 2}
\end{equation}%
The solution of this equation can be analytically written as%
\begin{equation}
\xi _{\pm }^{(2)}=C_{\pm }\exp \left[ i(k_{\pm }^{(2)}z-\omega t)\right]
+D_{\pm }\exp \left[ -i(k_{\pm }^{(2)}z+\omega t)\right] ,  \label{sol2}
\end{equation}%
where the wave number $k_{\pm }^{(2)}$ with each mode in the plasma is
defined as%
\begin{equation}
k_{\pm }^{(2)}\equiv \dfrac{1}{c}\sqrt{\sigma _{\pm }\left[ \sigma _{\pm
}\pm (m+2)\Omega \right] }.  \label{wave number 2}
\end{equation}%
\ 

In region (3) with pure vacuum $(z>L),$ the wave equation becomes%
\begin{equation}
\dfrac{d^{2}\xi ^{(3)}}{dz^{2}}+\dfrac{\omega ^{2}}{c^{2}}\xi ^{(3)}=0.
\label{wave eq 3}
\end{equation}%
Owing to the absence of rotating matter, two helical states satisfy the same
equation. We here dropped the notation `$\pm $'. The solution is easily
given by%
\begin{equation}
\xi ^{(3)}=E\exp \left[ i(k^{(3)}z-\omega t)\right] .  \label{sol3}
\end{equation}%
The wave number thus takes an ordinal form as 
\begin{equation}
k^{(3)}\equiv \dfrac{\omega }{c}
\end{equation}%
In the above expressions, $C_{\pm },D_{\pm }$ and $E$ are complex incidence,
reflection and transmission amplitudes in each region, respectively. These
wave amplitudes and the wave numbers determine the transmission rate of the
wave based on some boundary conditions. Mathematical treatment will be given
in section 2.4.

\subsection{Boundary Condition}

The physical property of Alfven waves changes at the bottom and at the top
of the rotating plasma layer, depending on the azimuthal wave number and the
angular velocity of the background. We now require boundary conditions in
the usual way in order to connect each wave solution continuously. From
equations (\ref{sol2}) and (\ref{sol3}), the continuity at the upper surface
of the plasma, $z=L,$ gives%
\begin{eqnarray}
C_{\pm } &=&E\dfrac{k_{\pm }^{(2)}+k^{(3)}}{2k_{\pm }^{(2)}}\exp \left[
-i\left( k_{\pm }^{(2)}-k^{(3)}\right) L\right] ,  \label{coeffi c} \\
D_{\pm } &=&E\dfrac{k_{\pm }^{(2)}-k^{(3)}}{2k_{\pm }^{(2)}}\exp \left[
i\left( k_{\pm }^{(2)}+k^{(3)}\right) L\right] .  \label{coeffi d}
\end{eqnarray}%
Substituting equations (\ref{coeffi c}) and (\ref{coeffi d}) into equation (%
\ref{sol2}) and then taking the derivative at the stellar surface, one
obtains%
\begin{equation}
\left. \dfrac{d}{dz}\ln \left( \xi _{\pm }^{(2)}\right) \right|
_{z=0}=ik_{\pm }^{(2)}\gamma ,  \label{bc at surf}
\end{equation}%
where $\gamma $ is a modification quantity due to background rotation of the
surrounding plasma defined as 
\begin{equation*}
\gamma =\dfrac{\varkappa _{+}\exp \left[ -i\varkappa _{-}L\right] -\varkappa
_{-}\exp \left[ i\varkappa _{+}L\right] }{\varkappa _{+}\exp \left[
-i\varkappa _{-}L\right] +\varkappa _{-}\exp \left[ i\varkappa _{+}L\right] }%
,
\end{equation*}%
with%
\begin{gather}
\varkappa _{+}=k_{\pm }^{(2)}+k^{(3)}, \\
\varkappa _{-}=k_{\pm }^{(2)}-k^{(3)}.
\end{gather}%
Detailed treatment and physical meaning of the $\gamma $ are given in the
Appendix.

In order to solve wave equation (\ref{wave eq 1}), we consider the complex
linear combination $\zeta =\zeta _{1}+i\zeta _{2}$ with specific solutions $%
\zeta _{1}$ and $\zeta _{2}$ to be satisfied with equation (\ref{bc at surf}%
). At the ends, one can obtain the explicit boundary condition at the
stellar surface $z=0,$ 
\begin{gather}
\left. \dfrac{d}{dz}\ln \left( \zeta _{1}+i\zeta _{2}\right) \right| _{z=0} 
\notag \\
=-k_{\pm }^{(2)}\left\{ \zeta _{1}(0)\Im \gamma +\zeta _{2}(0)\Re \gamma
\right\}  \notag \\
+ik_{\pm }^{(2)}\left\{ \zeta _{1}(0)\Re \gamma -\zeta _{2}(0)\Im \gamma
\right\} .
\end{gather}%
Logarithmic derivative of asymptotic solution has to be equal to that of
numerical solution in the deep interior $z\ll 0$, so that we request%
\begin{equation}
\left. \dfrac{d}{dz}\ln \left( \xi _{\pm }^{\text{asymp}}\right) \right|
_{z=-q}=\left. \dfrac{d}{dz}\ln \left( \zeta \right) \right| _{z=-q}.
\end{equation}%
This yields%
\begin{equation}
\dfrac{B_{\pm }}{A_{\pm }}=\dfrac{\left( \beta \left| z\right| ^{-1}+i\dfrac{%
d\psi _{\pm }}{dz}\right) \zeta +\dfrac{d\zeta }{dz}}{\left( -\beta \left|
z\right| ^{-1}+i\dfrac{d\psi _{\pm }}{dz}\right) \zeta -\dfrac{d\zeta }{dz}}%
\exp \left[ -2i\psi _{\pm }\right] .
\end{equation}%
Finally, the reflection $R_{\pm }$ and transmission coefficients $T_{\pm }$
of the waves with each helicity propagating from the crust through the
plasma toward the vacuum exterior are respectively expressed as%
\begin{equation}
R_{\pm }=\dfrac{\left| B_{\pm }\right| ^{2}}{\left| A_{\pm }\right| ^{2}},
\end{equation}%
\begin{equation}
T_{\pm }=\dfrac{k^{(3)}}{k_{\pm }^{(1)}}\dfrac{\left| E\right| ^{2}}{\left|
A_{\pm }\right| ^{2}}=1-R_{\pm }.  \label{trans rate}
\end{equation}

\section{Propagation}

\subsection{Dispersion Relation}

In this section, behavior of the torsional Alfven waves is discussed based
on the dispersion relations. Eikonal equation (\ref{eikonal}) depends on the
depth through the functions $h$ and $\widetilde{v}$. For strong magnetic
fields $B_{\text{o}}\geq 10^{14}$ G considered in this paper, it can be
shown that $\psi _{\pm }$ in equation (\ref{eikonal}) varies weakly with the
depth except for the inner region close to the star surface. The local
dispersion relation can be approximately written as 
\begin{equation}
k_{\pm }\approx \dfrac{1}{\widetilde{v}}\sqrt{\sigma _{\pm }\left[ \sigma
_{\pm }\pm \left\{ m(1-h)+2\right\} \Omega \right] }.
\label{dispersion relation}
\end{equation}%
This relation can also be obtained by taking a short wavelength limit,
namely, high frequency limit. The dimensionless function $h$ involved in
equation (\ref{dispersion relation}) varies from $h=1$ in the deep interior
to $h=0$ in the surface and exterior. This function represents the ratio
between the rest mass energy density and the effective magnetic energy
density. For $h=1$, the rest mass energy dominates over the magnetic energy.
This case corresponds to the classical limit. On the other hand, strong
magnetic energy $B_{\text{o}}\geq 10^{14}$ G easily overwhelms the rest mass
energy of the electrons or ions in the low density region, where $h\simeq 0$%
. In this case, relativistic displacement current should be included in the
analysis.

In order to investigate the propagation property of the wave, it is usefull
to define the phase velocity $v_{\text{p}}$ in a straightforward manner as 
\begin{equation}
v_{\text{p}}^{2}/c^{2}=\omega ^{2}/\left( ck_{\pm }\right) ^{2},
\label{phase velocity}
\end{equation}%
which is also related to the refraction index $N_{\pm }=c/v_{\text{p}}$ for
each mode. Note that $v_{\text{p}}^{2}$ is not necessarily positive. It is
well-known in plasma physics (e.g., Wolfgang and Rudolf 1996) that waves are
in general reflected at the cut-off points where $N_{\pm }\rightarrow 0$,
and are absorbed at the resonant absorption points where $N_{\pm
}\rightarrow \infty $. Since the frequency $\sigma _{\pm }$ in the rotating
frame is here confined within the positive regime $\sigma _{\pm }>0$, $k_{+}$
is always real, that is, neither cut-off nor resonant absorption points
appear in the positive helicity. On the other hand, there are both
absorption and cut-off points for the negative mode. In the remainder of
this section, our attention will be thus paid only to this mode for physical
interests. The dispersion relations with negative mode are schematically
shown for classical limit in Fig.2 and for relativistic limit in Fig.3,
respectively. In both diagrams the region $v_{\text{p}}^{2}/c^{2}<0$ stands
for the non-propagation, viz. evanescent zone. Vertical dotted line denotes
the cut-off frequency of the wave.

\begin{center}
(i) Classical Limit
\end{center}

As seen in Fig.2, evanescent zone appears only in the low frequency and
narrow band width $0<\sigma _{-}/\Omega <2$. In this classical treatment,
the cut-off of the wave appears at $\sigma _{-}/\Omega =$ $2$, which can be
physically interpreted as the fact that Coriolis force interrupts the wave
propagation in the rotating frame. In a high frequency region $\sigma
_{-}/\Omega >2$, waves are almost capable of propagating except for $\sigma
_{-}/\Omega =m$, at which waves are strongly absorbed by resonance with a
rotating background. This is because refractivity diverges\ in this
particular frequency. One notices that such a resonant absorption frequency $%
\sigma _{-}=m\Omega $ in our model is formally analogous to the
electron-cyclotron resonant frequency in the standard plasma physics
(Wolfgang and Rudolf 1996).

\begin{center}
(ii) Relativistic Limit
\end{center}

It is important to understand the new effect that appears in the
relativistic case. Figure 3 demonstrates that evanescent zone prevails in
the frequency regime $0<\sigma _{-}/\Omega <m+2$, whose band width is
therefore broadened by large $m$ modes. Cut-off frequency is given by $%
\sigma _{-}/\Omega =m+2$. It is only a high frequency wave with $\sigma
_{-}/\Omega >m+2$ that can always propagate in a WKB sense. By contrast with
the classical limit, resonant absorption $\sigma _{-}/\Omega =m$ does not
appear in this propagation regime, but in the evanescent regime.

\begin{figure}
\centering
\includegraphics[scale=0.8]{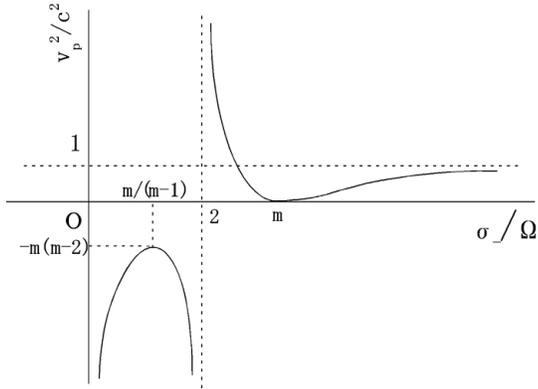}
\caption{
Dispersion relation for negative helicity waves with 
$m\geq 3$  in a classical limit, $h\rightarrow 1$.
The region below horizontal axis corresponds to the evanescent zone.
 Vertical dotted line denotes the cut-off frequency 
$\protect\sigma _{-}/\Omega =2$.%
The resonant absorption frequency is given by 
$\sigma _{-}/\Omega =2$.
}
\end{figure}
\begin{figure}
\centering
\includegraphics[scale=0.8]{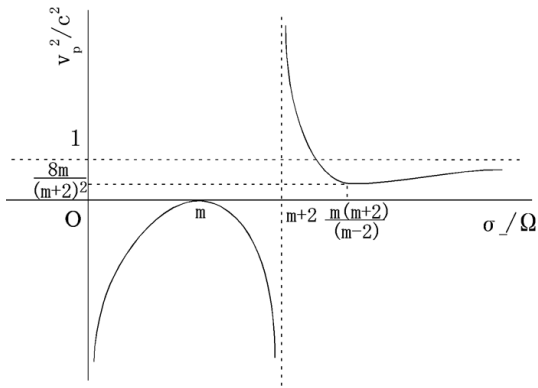}
\caption{
Same as Fig.2 in a relativistic limit, 
$h\rightarrow 0$. The cut-off
frequency is $\protect\sigma _{-}/\Omega =m+2$.
The resonant absorption occurs within an evanescent regime at
$\protect\sigma _{-}/\Omega =m.$
relation for negative helicity waves with $m\geq 3$
in a classical limit, $h\rightarrow 1$. 
The region below horizontal axis corresponds to the evanescent zone.
 Vertical dotted line
denotes the cut-off frequency $\protect\sigma _{-}/\Omega =2$.
The resonant absorption frequency is given by 
$\protect \sigma _{-}/\Omega =2$.
}
\end{figure}

\subsection{Structure of the Potential Barrier}

\begin{figure}
\centering
\includegraphics[scale=0.8]{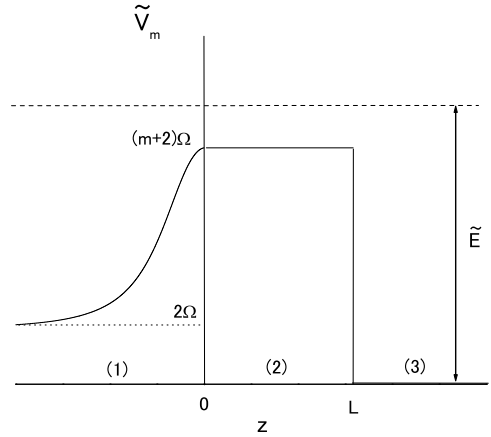}
\caption{
Spatial structure of the potential barrier for the negative mode 
as a function of 
distance. Horizontal dotted line denotes rotation frequency 
$\protect\sigma _{-}$ of the wave propagating above the potential.
}
\end{figure}
 As is obvious from Fig.3, evanescency can not be neglected in large $m$
mode. In this subsection, we turn to the investigation of the spatial
structure in the evanescent zone for large $m$. We now introduce a new
dependent variable $\Psi _{-}$ defined by $\Psi _{-}\equiv \widetilde{\mu }%
^{1/2}\xi _{-}$ for the amplitude of the negative mode with $\widetilde{z}%
\equiv \sqrt{\sigma _{-}}z$. Exploiting some arithmetic algebra, wave
equation ($\ref{wave eq 1}$) is now rewritten into a Sturm-Liouville type
differential equation 
\begin{equation}
\dfrac{d^{2}\Psi _{-}}{d\widetilde{z}^{2}}+\left[ \dfrac{1}{4\widetilde{\mu }%
^{2}}\left( \dfrac{d\widetilde{\mu }}{d\widetilde{z}}\right) ^{2}-\dfrac{1}{2%
\widetilde{\mu }}\dfrac{d^{2}\widetilde{\mu }}{d\widetilde{z}^{2}}+\dfrac{1}{%
\widetilde{v}^{2}}\left( \widetilde{E}-\widetilde{V}_{m}\right) \right] \Psi
_{-}=0,  \label{new wave eq}
\end{equation}%
where%
\begin{gather}
\widetilde{E}=\sigma _{-}, \\
\widetilde{V}_{m}=\left\{ m\left( 1-h\right) +2\right\} \Omega .
\end{gather}%
The first two terms concerning $\widetilde{\mu }$ in square brackets in
equation (\ref{new wave eq}) become very small for strong magnetic fields $%
B_{\text{o}}>10^{14}$ G, since $\widetilde{\mu }$ can be almost regarded as
a constant. One can compare equation (\ref{new wave eq}) with the
one-dimensional Schr\"{o}dinger equation for box-type potential in a
stationary state. Physical quantities $\widetilde{E}$ and $\widetilde{V}_{m}$
respectively correspond to \textit{wave energy} and \textit{potential} in a
formal sense. Strictly, potential $\widetilde{V}_{m}$ depends on the wave
energy $\widetilde{E}$ through the frequency $\sigma _{-}$. However, $\sigma
_{-}$ is here treated as a constant value irrespective of the position, so
that the following discussions are valid without loss of generality.

Schematic profile of the \textit{effective potential} $\widetilde{V}_{m}$
and the \textit{wave energy} $\widetilde{E}$ are given in Fig.4 as a
function of distance $z$. We draw the curve in region (1) somewhat
exaggeratedly. The \textit{potential} varies with the position through the
function $h$ and the background rotation $\Omega $. The \textit{potential}\
height is determined by the coupled quantity $\left( m+2\right) \Omega $ and
its width is given by the size $L$ of the corotating zone. This result means
that the large wave number lifts up the \textit{potential} barrier, only if
the background stellar medium rotates. In other words, if surrounding matter
is not dragged by the star, say, being in static state $\Omega =0$, then the 
\textit{potential} never rises, even though some high azimuthal waves exist.

Whether or not the wave can propagate and transmit out depends on the 
\textit{wave energy}, that is, wave frequency\ $\sigma _{-}$. Our argument
deserves to be specially emphasized in two explicit regimes; the evanescent
frequency mode $2\Omega <\sigma _{-}<(m+2)\Omega $ and the propagation mode $%
\sigma _{-}>(m+2)\Omega $.

\subsubsection{Evanescent Mode: $2\Omega <\protect\sigma _{-}<(m+2)\Omega $}

As is apparent in Fig.4, such low frequency waves excited in the interior
cannot help striking the potential barrier. Since refractivity is zero on
the critical curve $\sigma _{-}/\Omega =m\left( 1-h\right) +2$, outgoing
waves with $m\left( 1-h\right) +2<\sigma _{-}/\Omega <m+2$ are generally
reflected, when they reach the potential wall.

\subsubsection{Propagation Mode: $\protect\sigma _{-}>(m+2)\Omega $.}

In this case perturbation of the electromagnetic fields can propagate as a
wave. The present context in our model is almost concerned with the WKB
frequency range. The propagation sometimes exhibits a remarkable property,
if certain conditions are satisfied. As long as the wave frequency is much
greater than the \textit{potential} strength $\sigma _{-}\gg (m+2)\Omega $,
the \textit{potential} itself does not have much influence on the wave
behavior. However, if the frequency becomes commensurable to the \textit{%
potential} height $\sigma _{-}\gtrsim (m+2)\Omega $, the existence of a 
\textit{potential} barrier cannot be ignored. Especially if the wavelength
is comparable to the \textit{potential} width, i.e., $k_{-}L\sim 1$, then
the waves will interfere with the \textit{potential} barrier and their
behavior will be strongly altered. This inherent property is expected to be
more evident near the threshold frequency $\sigma _{-}\sim (m+2)\Omega $.
Motivated by this general consideration, we explicitly calculate the
transmission rate in the next section. The numerical parameters are adopted
to satisfy the above condition, $L\sim 1/k_{-}\sim 10^{6}$ cm and $m=\omega
_{\text{max}}/\Omega \sim 10^{6}$. Using these parameters, we will examine
how and to what extent the propagation and transmission are affected by the 
\textit{potential} barrier.

\section{Transmission}

\subsection{Numerical Calculation}

We numerically calculate the transmission rates (\ref{trans rate}) of the
wave propagating from the deep crust, through the corotating plasma
envelope, toward the vacuum exterior subject to matching conditions at each
boundaries. As already mentioned above, only the negative helicity wave is
intriguing for physical interests. Evanescent modes in a low frequency
regime $0<\sigma _{-}<\left\{ m(1-h)+2\right\} \Omega $ are excluded from
this calculation. This constraint on wave frequency thereby guarantees that
all waves are capable of propagating in a WKB sense. One can roughly
estimate a typical wave frequency $\omega $ measured in the inertial frame
by approximating as $\omega \sim \widetilde{v}/q$. Appropriate frequency
thus lies in a finite range $10^{3}\lesssim \omega \lesssim 10^{6}$ s$^{-1}$.

The objective of this section is to explore the effects of azimuthal wave
number $m$ and the size $L$ of the corotating plasma medium on the wave
transmission. We here consider two kinds of specific torsional waves whose
azimuthal number is; (i) the fundamental, $m=1$ and (ii) much larger than
one, $m=\omega _{\text{max}}/\Omega =10^{6}$. In each case, we further
consider two apparently different circumstances in the exterior $L=R\sim
\lambda _{-}\sim 10^{6}$ cm and in the absence of plasma $L=0$ cm.

Figures 5 and 6 respectively demonstrate the transmission coefficients for
the negative helicity waves with two extreme cases (i) $m=1$ and (ii) $%
m=10^{6}$ as a function of the wave frequency measured in the inertial frame
when $B_{\text{o}}=10^{15}$ G, typical for magnetars. In both figures, solid
and dotted lines denote the results of $L=10^{6}$ and $L=0$ cm,
respectively. The surrounding plasma is here assumed to corotate with the
same angular velocity as that of the star, $\Omega =1$ s$^{-1}$. In each
case, we got the following results.

\begin{center}
(i) The Fundamental Wave Number: $m=1$
\end{center}

As seen in Fig.5, the transmission curve in the case of $L=10^{6}$ cm
slightly shows a \textit{wiggling} behavior. This curve intersects that of $%
L=0$ at $\omega \sim 10^{4}$ s$^{-1}$. However, the difference between both
cases becomes undistinguishable in the high frequency regime $\omega \gtrsim
10^{5}$ s$^{-1}$, since the wave frequency is much higher than the critical
one $\left( m+2\right) \Omega =3$ s$^{-1}$, which appeared as the potential
barrier in section 3. As the wave frequency gets higher, the transmission
rate approaches unity asymptotically. Since the overall property does not
depend on $L$, the potential barrier due to rotating plasma does not make
any significant influence on the wave transmission for small azimuthal
number $m\sim 1$.

\begin{center}
(ii) Highly Azimuthal Wave Number: $m=10^{6}$
\end{center}

In this high $m$ mode, we can find some suprising results. Figure 6 shows
that the transmission rate of $L=10^{6}$ cm is drastically enhanced at some
selected frequencies. More importantly, such enhancements occur periodically
at specific frequencies. At the top of the first wing $(n=1)$ $\omega _{%
\text{t}}=2.2\times 10^{3}$ s$^{-1}$, the rate reaches the maximum $T_{\text{%
max}}$=$0.71$, which corresponds to approximately $70$ times larger than
that of the first bottom $\omega _{\text{b}}=7.0\times 10^{3}$ s$^{-1}$.
Comparing with $L=0$ case at the same frequency $\omega _{\text{t}}$, this
maximum $T_{\text{max}}$ amounts to $230$ times larger. In this way, plasma
effect is important at each top frequency, but unimportant at any bottom
frequency. As shown in Fig.6, the transmission coefficient through the
plasma layer is exactly equal to that of $L=0$ case at arbitrary wing
bottom. In general rotating plasma has an effect in helping the escape of
the waves except for the bottom frequencies. This point is different from
that of case (i).

\begin{figure}
\centering
\includegraphics[scale=0.8]{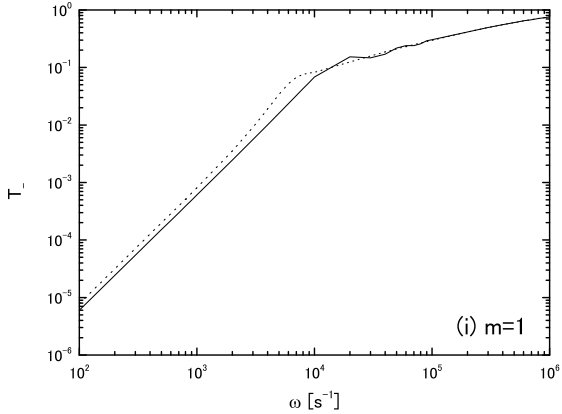}
\caption{
 Transmission coefficients of the negative helicity wave with
the fundamental mode $m=1$%
 as a function of frequency when $B_{\text{o}}=10^{15}$ G and 
$\Omega =1$ s$^{-1}$.%
Solid and dotted lines denote the results of $L=10^{6}$%
and $L=0$cm, respectively.
}
\end{figure}
\begin{figure}
\centering
\includegraphics[scale=0.8]{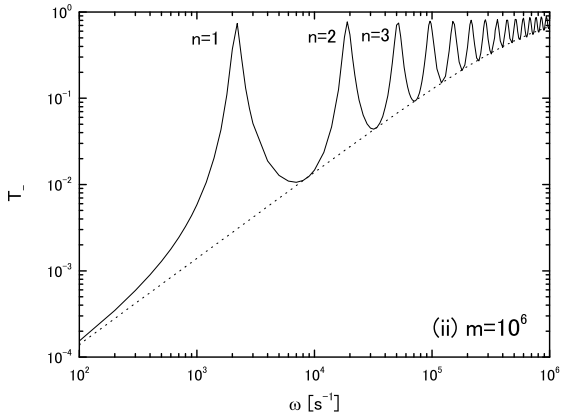}
\caption{
Same as in Figure
5, but for high azimuthal number $m=10^{6}$. In this high 
$m$ mode transmission rate is highly enhanced at some
selected frequencies due to the plasma envelope. 
}
\end{figure}
%
 In order to understand the transmission enhancement, we have also
numerically computed the integral $\int Td\omega $ in some frequency bands
for both $L=10^{6}$ and $L=0$ cm. We obtain some $13$ times enhancement in
the narrow band $2\Omega <\omega <\omega _{\text{b}}$ and $1.4$ times in the
broad band $2\Omega <\omega <\omega _{\text{max}}$ compared to the results
of $L=0$. Such enhancement and periodic variation slowly decrease with
increasing wave frequency. This property almost vanishes when the frequency
becomes comparable to the critical frequency $\left( m+2\right) \Omega \sim
10^{6}$ s$^{-1}$.

Periodic enhancements in our numerical calculations can be interpreted as a
consequence of wave interference within the rotating plasma cavity. In the
following subsection, we investigate our results more quantitatively in a
simplified analytical method.

\subsection{Analytical Calculation}

Periodic enhancements on the wave transmission in our model can be
satisfactorily rationalized by comparing with the wave propagation in
homogeneous multi-media. Non-relativistic Alfven resonance itself in the
magnetosphere around the earth and the sun have been widely recognized and
discussed theoretically and observationally (for recent reviews see
Leonovich and Mazur 1997; Waters 2000; Li and Wang 2001). Hollweg (1983) has
theoretically studied WKB wave propagation in three homogeneous layers
labeled by (1), (2) and (3) separated at $z=0$ and $z=L$ as shown in Fig.7.
In his work Alfven waves are simply assumed to have constant wave numbers $%
k^{(1)}$, $k^{(2)}$ and $k^{(3)}$ in each region. In this homogeneous model,
the transmission coefficient $T$ can be analytically calculated by taking
boundary conditions at the discontinuities $z=0$ and $z=L$, 
\begin{gather}
T=\dfrac{4k^{(3)}}{k^{(1)}}\left[ \left\{ 1+\dfrac{k^{(3)}}{k^{(1)}}\right\}
^{2}\cos ^{2}(k^{(2)}L)\right.  \notag \\
\left. +\left\{ \dfrac{k^{(3)}}{k^{(2)}}+\dfrac{k^{(2)}}{k^{(1)}}\right\}
^{2}\sin ^{2}(k^{(2)}L)\right] ^{-1}.  \label{trans2}
\end{gather}%
\ 

Although the physical situation is different from our present work, similar
expressions could be found also in our model. This formula (\ref{trans2})
implies that the transmission rate has a periodic structure depending on the
wave number $k^{(2)}$ and the characteristic scale $L$ of the intermediate
layer unless $k^{(2)}L\sim 0$. Extrema of the transmission rate work out to
be 
\begin{equation}
T_{\text{max}}=\left\{ 
\begin{array}{c}
\dfrac{4k^{(1)}k^{(3)}}{\left( k^{(1)}+k^{(3)}\right) ^{2}}\text{ \ \ at }%
k^{(2)}L=n\pi , \\ 
\dfrac{4k^{(1)}k^{(3)}/(k^{(2)})^{2}}{\left(
1+k^{(1)}k^{(3)}/(k^{(2)})^{2}\right) ^{2}}\text{ \ \ at }k^{(2)}L=\left(
2n-1\right) \pi /2,%
\end{array}%
\right.
\end{equation}%
with $n=1,2,\cdots $. Providing that the wave number in each region
satisfies the inequalities 
\begin{gather}
k^{(1)}\gg k^{(2)}\gg k^{(3)},  \label{ineq k} \\
k^{(2)}L\gtrsim 1,  \label{ineq L}
\end{gather}%
the coefficient of $\cos ^{2}(k^{(2)}L)$ in equation (\ref{trans2}) becomes
dominant compared to that of $\sin ^{2}(k^{(2)}L)$. In this limit, equation (%
\ref{trans2}) can be well approximated by%
\begin{equation}
T\approx \dfrac{4k^{(1)}k^{(3)}}{\left( k^{(1)}+k^{(3)}\right) ^{2}}\cos
^{-2}(k^{(2)}L).  \label{trans 2}
\end{equation}%
One notices that if $k^{(1)}\sim k^{(2)}\sim k^{(3)}$, the transmission
shows neither periodic variations nor enhancements. Consequently,
inequalities (\ref{ineq k}) and (\ref{ineq L}) give a set of resonant
conditions of the wave. Sterling and Hollweg (1984) have subsequently
considered a three layer model for solar flare, which is composed of a
chromosphere, a spicule and a corona. In that work they have confidently
suggested the possibility of Alfvenic resonance on solar spicules and have
shown a new aspect of spicules which may account for occasionally twisting
motions of magnetic field lines, even when above resonant conditions
approximately hold. Conditions (46) and (47) may hold true also in our model
except close to the thin regime beneath the stellar surface, whenever large
twisted waves with $m\gg 1$ propagate in the rotating background.

We can now quantify the particular frequencies at which the wave resonance
occur. By solving the quadratic equation $\left( \sigma _{-}\right)
^{2}-\left( m+2\right) \Omega \sigma _{-}-\left( k_{-}^{(2)}\widetilde{v}%
\right) ^{2}=0$, together with the periodic condition $k_{-}^{(2)}L=\left(
2n-1\right) \pi $, the eigen frequencies of the negative mode are obtained as%
\begin{equation}
\omega _{m,n}=\dfrac{\left( 2-m\right) }{2}\Omega +\sqrt{\left( \dfrac{m+2}{2%
}\Omega \right) ^{2}+\left( \dfrac{\left( 2n-1\right) c\pi }{2L}\right) ^{2}}%
,  \label{resonant freq}
\end{equation}%
with $m,n=1,2,3,\cdots $. This formula shows that the resonant frequencies
depend critically on the azimuthal number $m$ and the plasma cavity length $%
L $. They gently decrease with an increase in $L$ or $m$. Physically, this
means that it takes longer time for the wave to go back and forth between
the stellar surface and the top of the plasma layer and then this wave
interferes with another one propagating from the crust into the cavity. In
extremal cases, one finds 
\begin{equation}
\lim\limits_{L\rightarrow \infty }\omega _{m,n}=\lim\limits_{m\rightarrow
\infty }\omega _{m,n}=2\Omega ,
\end{equation}%
which coincides with the cut-off frequency in the classical limit. At the
same time, transmission peaks become blended with neighboring resonance,%
\begin{equation}
\lim\limits_{L\rightarrow \infty }\Delta \omega
_{m}=\lim\limits_{m\rightarrow \infty }\Delta \omega _{m}=0,
\end{equation}%
where $\Delta \omega _{m}\equiv \omega _{m,n}-\omega _{m,n-1}$.

For $\Omega =1$ s$^{-1}$ and $L=10^{6}$ cm, from equation (\ref{resonant
freq}) one can calculate some representative resonant frequencies; the
fundamental mode frequency $\omega _{10^{6},1}=8.8\times 10^{3}$ s$^{-1}$,
the second mode $\omega _{10^{6},2}=7.4\times 10^{4}$ s$^{-1}$ and the third 
$\omega _{10^{6},3}=1.9\times 10^{5}$ s$^{-1}$. Our numerical work also
gives similar results; $\omega _{10^{6},1}=2.2\times 10^{3}$ s$^{-1}$, $%
\omega _{10^{6},2}=1.9\times 10^{4}$ s$^{-1}$ and $\omega
_{10^{6},3}=5.2\times 10^{4}$ s$^{-1}$, respectively. Our results exhibit
slightly positive deviations from the analytical ones. Recall that the wave
number $k^{(1)}$ in our magnetar model cannot be regarded as a constant,
since the mass density drastically changes in the vicinity of the stellar
surface. This discrepancy would be probably attributed to the inhomogeneity
in the crust. If a sharp boundary is formed at the stellar surface, such a
difference may probably become small. From a comparison with the analytic
model, we concluded that the transmission enhancements obtained in our model
are thought to be a result of the Alfven resonance on the rotating plasma
cavity due to the potential barrier.
%
\begin{figure}
\centering
\includegraphics[scale=1.0]{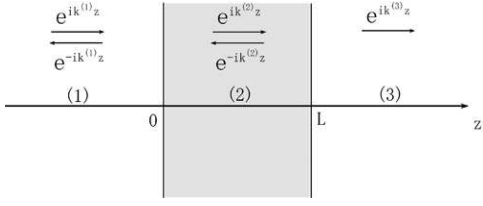}
\caption{
 Three homogeneous model separated by two
discontinuities at $z=0$ and $z=L$. If
periodic boundary condition $k^{(2)}L=n\protect\pi $
$(n=1,2,\cdots )$ holds, intermediate layer acts as a
resonant cavity and transmission is highly enhanced.
}
\end{figure}

\section{Electromagnetic Field Structure}

In the preceding section we have elucidated that the transmission rates of
wave are highly enhanced at some selected frequencies owing to the resonance
effect in the rotating plasma cavity. Resonance in the plasma portion has
another spectacular nature. In this section, it will be shown that resonance
has a great impact not only on the transmission rates, but also on the
electromagnetic field structure associated with Alfven waves. Most of our
applications will once again concern the WKB frequency regime of the
negative helicity $\sigma _{-}>\left\{ m(1-h)+2\right\} \Omega $. Hereafter
we omit the subscript `$-$' for simplicity.

By substituting equation (\ref{sol2}) into (\ref{dE2}) and (\ref{dB2}), we
can explicitly have the expressions for electromagnetic field amplitudes in
the plasma zone in terms of $k^{(i)}$, normalized by a local magnetic field $%
B_{\text{o}}$%
\begin{gather}
\dfrac{\left| \delta B_{\varpi }\right| ^{2}}{B_{\text{o}}^{2}}=\dfrac{%
\left| \delta B_{\phi }\right| ^{2}}{B_{\text{o}}^{2}}=\dfrac{1}{\Omega
^{2}\varpi ^{2}}\dfrac{\left| c\delta E_{z}\right| ^{2}}{B_{\text{o}}^{2}} 
\notag \\
=f_{m}(\varpi )\left| \dfrac{d\xi ^{(2)}}{dz}\right| ^{2}  \notag \\
=f_{m}(\varpi )\dfrac{k_{m}^{(1)}}{k^{(3)}}T_{m}\left| A\right| ^{2}  \notag
\\
\times \left[ (k_{m}^{(2)})^{2}\sin ^{2}k_{m}^{(2)}\left( z-L\right)
+(k^{(3)})^{2}\cos ^{2}k_{m}^{(2)}\left( z-L\right) \right] ,
\end{gather}%
and%
\begin{gather}
\dfrac{\left| c\delta E_{\varpi }\right| ^{2}}{B_{\text{o}}^{2}}=\dfrac{%
\left| c\delta E_{\phi }\right| ^{2}}{B_{\text{o}}^{2}}=f_{m}(\varpi )\sigma
_{m}^{2}\left| \xi ^{(2)}\right| ^{2}  \notag \\
=f_{m}(\varpi )\left( \sigma _{m}\right) ^{2}\dfrac{k_{m}^{(1)}}{%
(k_{m}^{(2)})^{2}k^{(3)}}\left| A\right| ^{2}T_{m}\times  \notag \\
\left[ (k_{m}^{(2)})^{2}\cos ^{2}k_{m}^{(2)}(z-L)+(k^{(3)})^{2}\sin
^{2}k_{m}^{(2)}(z-L)\right] ,
\end{gather}%
where $f_{m}(\varpi )$ is a radial profile given by $f_{m}(\varpi )\equiv
(\varpi /\varpi _{\text{pc}})^{2(m-1)}$ $(m=1,2,\cdots )$ and the other
notations have the same meaning as the previous ones. The longitudinal
component of the perturbed magnetic fields is always zero $\delta B_{z}=0$,
because the matter has been assumed to be vertically immobile in the present
work. On the contrary, only if the background rotates, the electric fields
have a longitudinal component whose structure is essentially identical with
that of the magnetic fields. Thereby we only have to investigate the
transverse components $(\varpi ,\phi )$ of the fields.

For the sake of exploring the dependence of the azimuthal number $m$ on the
field structure, we once more restrict our discussions within two kinds of
extreme cases: (i) the fundamental mode $m=1$ and (ii) the large azimuthal
wave number $m=l\left( \gg 1\right) $. In this limit, formulae (49) and (50)
can be well approximated by%
\begin{gather}
\dfrac{\left| \delta B_{\varpi }\right| ^{2}}{B_{\text{o}}^{2}}=\dfrac{%
\left| \delta B_{\phi }\right| ^{2}}{B_{\text{o}}^{2}}=\dfrac{1}{\Omega
^{2}\varpi ^{2}}\dfrac{\left| c\delta E_{z}\right| ^{2}}{B_{\text{o}}^{2}} 
\notag \\
=\left\{ 
\begin{array}{c}
f_{m}(\varpi )k_{m}^{(1)}k^{(3)}T_{m}\left| A\right| ^{2}\text{ \ \ for }m=1,
\\ 
f_{m}(\varpi )k_{m}^{(1)}k^{(3)}\left( \dfrac{k_{m}^{(2)}}{k^{(3)}}\right)
^{2}T_{m}\left| A\right| ^{2}\sin ^{2}k_{m}^{(2)}\left( z-L\right) \text{ \ }
\\ 
\text{for }m=l,%
\end{array}%
\right.  \label{deruta B2}
\end{gather}%
and%
\begin{align}
\dfrac{\left| c\delta E_{\varpi }\right| ^{2}}{B_{\text{o}}^{2}}& =\dfrac{%
\left| c\delta E_{\phi }\right| ^{2}}{B_{\text{o}}^{2}}  \notag \\
& =\left\{ 
\begin{array}{c}
f_{m}(\varpi )\sigma _{m}^{2}\dfrac{k_{m}^{(1)}}{k^{(3)}}T_{m}\left|
A\right| ^{2}\text{ \ \ for }m=1, \\ 
f_{m}(\varpi )\sigma _{m}^{2}\dfrac{k_{m}^{(1)}}{k^{(3)}}T_{m}\left|
A\right| ^{2}\cos ^{2}k_{m}^{(2)}\left( z-L\right) \text{ }\  \\ 
\text{for }m=l.%
\end{array}%
\right.  \label{deruta E2}
\end{align}%
Here we have dropped some small terms coupled to $(k^{(3)})^{2}$, since
inequality $k_{m}^{(1)}\gg k_{m}^{(2)}\gg k^{(3)}$ holds for large $m$.
Equations (\ref{deruta B2}) and (\ref{deruta E2}) have the implication that
the fields are almost constant for small $m$ modes, but are sinusoidally
changed with $z$ for large $m$.

In principle, the absolute value of the incident wave amplitude $\left|
A\right| $ cannot be determined in our linearized theory. If we are allowed
to assume that $f_{m=l}\left( \varpi \right) \left| A\right| _{m=l}^{2}\sim
f_{m=1}\left( \varpi \right) \left| A\right| _{m=1}^{2}$, then the ratios of
electromagnetic field amlitude with $m=l$ to those with $m=1$ are
approximately given by%
\begin{equation}
\dfrac{\left| \delta B_{i}\right| _{m=l}^{2}}{\left| \delta B_{i}\right|
_{m=1}^{2}}\approx \left( 1+\dfrac{l\Omega }{\omega }\right) ^{2}\left( 1-%
\dfrac{2\Omega }{\omega }\right) \dfrac{T_{m=l}}{T_{m=1}}\sin
^{2}k_{m}^{(2)}\left( z-L\right) ,
\end{equation}%
and 
\begin{equation}
\dfrac{\left| \delta E_{i}\right| _{m=l}^{2}}{\left| \delta E_{i}\right|
_{m=1}^{2}}\approx \left( 1+\dfrac{l\Omega }{\omega }\right) ^{3}\left( 1+%
\dfrac{\Omega }{\omega }\right) ^{-2}\dfrac{T_{m=l}}{T_{m=1}}\cos
^{2}k_{m}^{(2)}\left( z-L\right) ,
\end{equation}%
with $i=\varpi ,\phi $. These quantities are much larger than unity because
of the extra factor $\left( 1+l\Omega /\omega \right) $ except for some
special locations $z=n\pi /k_{m}^{(2)}$ and $z=(2n-1)\pi /2k_{m}^{(2)}$ $%
(m,n=1,2,\cdots ),$ which correspond to the nodes of the standing Alfven
wave.
%
\begin{figure}
\centering
\includegraphics[scale=0.8]{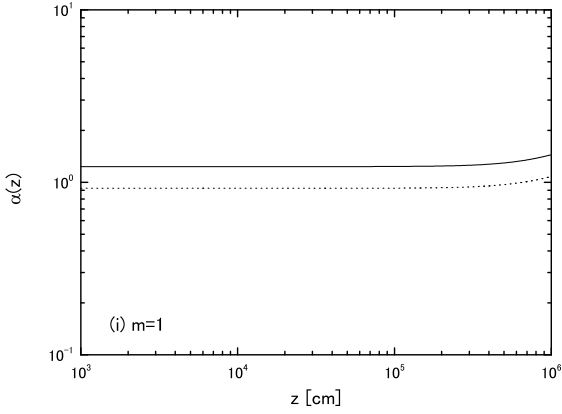}
\caption{
Spatial structure of perturbed magnetic field with 
$m=1$ in the
rotating plasma region normalized by that in the absence of plasma at some
resonant frequencies in the first wing. The solid and the dotted lines
denote the result at the top of wing 
$\protect\omega _{\text{t}}=2.2\times10^{3}$ s$^{-1}$
and at the bottom 
$ \protect\omega _{\text{b}}=7.0\times 10^{3}$  s$^{-1}$, respectively. 
}
\end{figure}
\begin{figure}
\centering
\includegraphics[scale=0.8]{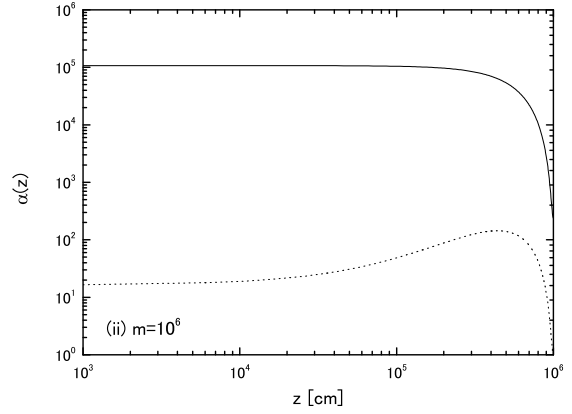}
\caption{
Same as in Fig.8 for $m=10^{6}$. 
Field amplitudes are drastically and
periodically enhanced for large $m$. This enlargement is
attributed to the resonance transmission and the conservation of energy flux
as a multiplier effect. 
}
\end{figure}

We will now examine the effect of the resonance in the rotating plasma
cavity, whose thickness is given by $L\sim R=10^{6}$ cm, on the spatial
structure of the perturbed electromagnetic fields. Let us designate by $%
\alpha $ the amplitude of the perturbed magnetic field within the plasma\
normalized by that in the absence of the plasma 
\begin{eqnarray}
\alpha (z) &\equiv &\dfrac{\left| \delta B_{i}\right| _{L=R}^{2}}{\left|
\delta B_{i}\right| _{L=0}^{2}}  \notag \\
&\approx &\dfrac{T_{L=R}}{T_{L=0}}\dfrac{c^{2}}{\omega ^{2}}\times  \notag \\
&&\left[ \left\{ (k_{m}^{(2)})^{2}-\dfrac{\omega ^{2}}{c^{2}}\right\} \sin
^{2}k_{m}^{(2)}\left( z-L\right) +\dfrac{\omega ^{2}}{c^{2}}\right] .\text{
\ \ }  \label{alpha}
\end{eqnarray}%
The first term in square brackets is related to the deviation due to high
torsional modes in the rotating background from the dispersion relation in a
vacuum. The ratio $\alpha $ is plotted against the height from the stellar
surface for (i) $m=1$ in Fig.8 and (ii) $m=10^{6}$ in Fig.9, respectively.
Solid lines denote the numerical results at the top of the wing $\omega _{%
\text{t}}=2.2\times 10^{3}$ s$^{-1}$, while dotted lines denote the bottom $%
\omega _{\text{b}}=7.0\times 10^{3}$ s $^{-1}$. The transmission rates at
resonant frequencies obtained in section 3 are appropriately used in this
calculation.

\begin{center}
(i) The Fundamental Wave Number: $m=1$
\end{center}

As shown in Fig.8, in the case of $m=1,$ the wave amplitude keeps one order
of magnitude in the plasma. This result agrees well with the fact that the
dispersion relation for small $m\sim 1$ is almost equal to that of the
electromagnetic fields in pure vacuum, $(k_{m}^{(2)})^{2}\sim \omega
^{2}/c^{2}$. The background rotation therefore does not affect this small $m$
modes.

\begin{center}
(ii) Highly Azimuthal Wave Number: $m=10^{6}$
\end{center}

Some significant differences can be found in this high mode. The perturbed
magnetic field strength $\alpha $ for $m=10^{6}$ at the resonant frequency $%
\omega _{\text{t}}$ is drastically amplified up to $\alpha \sim 10^{5}$ at $%
z\lesssim 10^{5}$ cm. Then the field strength $\alpha $ is fallen down to $%
\alpha =T_{L=R}/T_{L=0}\sim 230$ at the top of the plasma layer, in which
the first node of the standing Alfven wave is formed. Even at the bottom
frequency $\omega _{\text{b}}$ the strength $\alpha $ has about $10^{2}$ at $%
z\simeq 5\times 10^{5}$ cm.

In the same way, the normalized amplitude of the perturbed electric field is
expressed by%
\begin{eqnarray}
\beta (z) &\equiv &\dfrac{\left| \delta E_{i}\right| _{L=R}^{2}}{\left|
\delta E_{i}\right| _{L=0}^{2}}  \notag \\
&=&\dfrac{T_{L=R}}{T_{L=0}}\left( 1+\dfrac{m\Omega }{\omega }\right) ^{2}%
\dfrac{1}{(k_{m}^{(2)})^{2}}\times  \notag \\
&&\left[ \left\{ (k_{m}^{(2)})^{2}-\dfrac{\omega ^{2}}{c^{2}}\right\} \cos
^{2}k_{m}^{(2)}\left( z-L\right) +\dfrac{\omega ^{2}}{c^{2}}\right] ,\ 
\label{beta}
\end{eqnarray}%
which is plotted in Figs.10 and 11. When $m=10^{6}$, the field amplitude $%
\beta $ of $\omega _{\text{t}}$ has almost the same magnitude $10^{5}$ as $%
\alpha $ near the stellar surface $z\lesssim 10^{4}$ cm. But near the top of
the plasma $z\sim 10^{6}$ cm, in contrast to $\alpha $, $\beta $ has a
maximum $5\times 10^{7}$ which corresponds to the loops of the standing
wave. Approximately, $\beta $ is much greater than $\alpha $ because of the
extra term $\left( 1+m\Omega /\omega \right) ^{2}\sim 10^{6}$ in equation (%
\ref{beta}).

Such a field amplification can also be explained by considering a
conservation of energy flux $F\sim \widetilde{\rho }v\omega \left| \xi
\right| ^{2}$. The phase velocity $v$ of the mode $m\gg 1$ in the plasma
atmosphere is much smaller than that in the vacuum, as is confirmed in
equations (\ref{dispersion relation}) and (\ref{phase velocity}). Assuming
the same value $\left| \xi \right| $ at the stellar surface, which is
irrelevant to $L$, the amplitude $\left| \xi \right| (\varpropto v^{-1/2})$
is enhanced in the rotating plasma region so as to compensate for the
slowing down of the wave propagation.
%

\begin{figure}
\centering
\includegraphics[scale=0.8]{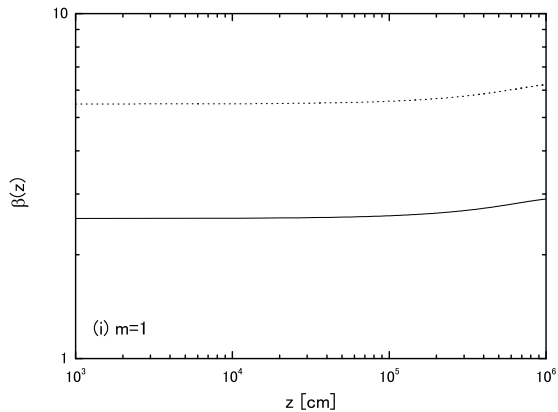}
\caption{
Perturbed electric fields with $m=1$
in the rotating plasma at resonant frequencies in the first wing.
Solid line and dotted line respectively correspond to 
$\protect\omega _{\text{t}}=2.2\times 10^{3}$
s$^{-1}$ and 
$\protect\omega _{\text{b}}=7.0\times 10^{3}$ s$^{-1}$.
}
\end{figure}
\begin{figure}
\centering
\includegraphics[scale=0.8]{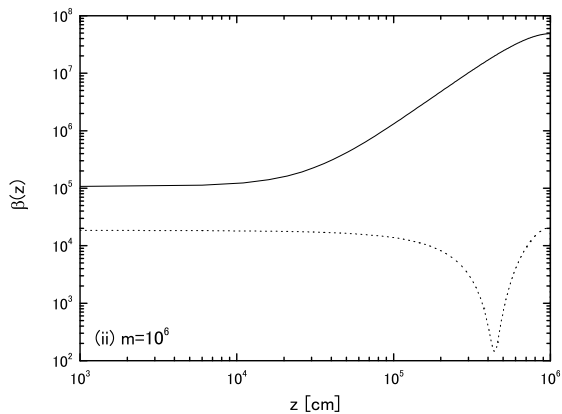}
\caption{
Same as in Fig.10 for $m=10^{6}.$
}
\end{figure}

\section{Summary and Discussion}

We have studied the propagation and transmission of torsional Alfven waves
along the rotation axis on magnetars by using three-layered cylindrical
model. If the intermediate plasma layer rotates with angular velocity $%
\Omega $, the propagation property for largely twisted waves with $m>\omega
/\Omega $ is drastically modified. This middle zone can be regarded as a
kind of a potential barrier, whose height is explicitly specified by $%
(m+2)\Omega $. Waves having an angular velocity of $2\Omega $ originates
from Coriolis force in the classical limit, while an additional quantity $%
m\Omega $ comes from the relativistic effect. This therefore means that the
potential is lifted up higher in the large $m$ modes on the rotating
background. It should be emphasized that this new finding is purely
attributed to having incorporated the displacement current into the model.

We have numerically computed the transmission rates of the torsional waves
with large value $m\gg 1$ driven in the crust, through the corotating plasma
with a finite size $L\sim R$, into the exterior. We found that the
transmissions are strongly enhanced for large $m\approx \omega /\Omega $ at
selected wave frequencies satisfied with a periodic condition $kL\simeq
\left( 2n-1\right) \pi /2$ $(n=1,2,\cdots )$. Such a transmission
enhancement arises because the rotating plasma forms a resonant cavity which
traps the wave energy by virtue of the strong reflections occuring at the
transition region at the bottom and at the top of the plasma layer.
Efficient transmissions due to resonance is also compatible with the fact
that the reflection of waves at the stellar surface can almost completely be
extinguished. Magnetospheric waveguide or resonance cavity can thus actually
generate a set of coherent eigenmodes for high $m$. Resonance may be a
signature of fundamental processes by which high order torsional
oscillations of waves, if there exists, can transfer much more energy out of
the magnetars especially in the low frequency regime.

It has been well-known in quantum mechanics that when the potential width is
comparable to (or integer times) de Broglie wavelength of electrons $L\sim
n\hbar /m_{\text{e}}v$, incident wave flux is bounded by finely tuning a
phase relation with inversely propagating waves (e.g., Schiff 1949). Then
resonance is excited in the potential zone and the wave flux spends much of
its time in the resonance cavity. Similar resonance may occur also in our
macroscopic model, if the characteristic scale of the evanescent zone is
comparable to the wavelength of the Alfven wave.

We have investigated the spatial structure of the perturbed electromagnetic
fields propagating in the rotating plasma, when standing Alfven waves are
formed. Resonance in the plasma cavity for high $m$ modes has another
spectacular aspect. At some points magnetic fields of the Alfven waves are
drastically amplified even up to a few $10^{5}$ times as large as those in
the absence of plasma or in the static exterior. This amplification can be
straightforwardly explained by both the transmission enhancement due to
resonance and the conservation of WKB energy flux.

In this way, once the Alfven waves are resonantly excited and transmissions
are greatly enhanced, a substantial part of the wave energy will probably be
transferred into the ambient charged particles such as electrons or
positrons confined within the fire balls associated with the bursts on
magnetars. The charged particles entwined around magnetic field lines will
then be violently swayed. These disturbances may eventually produce high
energy $\gamma $-ray or X-ray emissions from the magnetars. More
importantly, if magnetic field lines at some altitudes are strongly
distorted from equilibrium state by resonance and plasma fluid contracts
into the surface, the field lines with antiparallel components will approach
each other. Such geometry of field lines will give rise to possible magnetic
reconnection type events as Thompson and Duncan (2000) have suggested.

The behavior of resonant MHD waves within the corona or spicules erupted
from the solar surface have been studied so far. In fact, field line
resonance has been found to occur on magnetic shells in the magnetosphere of
the earth. But their physical treatment is inevitably restricted to the very
weak magnetic fields. Resonant property found in our magnetar model is very
similar to that in solar astrophysics or planetary physics. We can expect
that the same mechanism works also on the relativistic Alfven waves even in
a different environment, that is, in an extreme circumstance accompanied by
very strong magnetic fields such as the magnetars.

As a final remark, we shall address on the angular velocity $\Omega $ of the
star and the azimuthal wave number $m$. In our model, the rotating
background is throughout assumed to have a slow angular velocity $\Omega =1$
s$^{-1}$, which is actually observed in SGRs and AXPs. We concentrate our
treatment on the torsional waves with a specific value $m=10^{6}$ as a
possible limit of high $m$. As shown in our work, the striking behavior of
the resonance necessarily requires high order torsional modes, $m>\omega /(1$
s$^{-1})\simeq 10^{3}$-$10^{6}$. It is not clear whether or not such modes
with high wave number $m$ realistically exist on the magnetars. However, one
should keep it in mind that the local dispersion relation of the Alfven
waves are almost determined by the coupled quantity $m\Omega $ except for
the high frequency regime. This means that the physical behavior of the wave
with high $m=10^{3}$-$10^{6}$ on the slowly rotating background$\ \Omega =1$
s$^{-1}$ is essentially equivalent to that of small $m=1$-$10^{3}$ and fast
rotation $\Omega =10^{3}$ s$^{-1}$. Neutron stars are often expected to have
been a rapid rotator with $\Omega \sim 10^{3}$ s$^{-1}$ at a star-born
period. Our results would thus become much more effective especially on
young magnetars.\bigskip

\begin{center}
{\Large Appendix}
\end{center}

The physical effects of $\gamma $ can be clarified by taking some explicit
limits. Specifically if one ignores the substantial thickness of the plasma
layer $L\rightarrow 0$, then the wave number of region (2) should be
replaced with that of region (3), $k_{\pm }^{(2)}\rightarrow k^{(3)}$. This
yields that $\varkappa _{+}\rightarrow 2k^{(3)},\varkappa _{-}\rightarrow 0$
and $\gamma \rightarrow 1$. In this limit, the boundary condition reduces to 
$d/dz[\ln \xi _{\pm }^{(2)}]=ik^{(3)}$, which clearly corresponds to the
simple case that the exterior of the star is filled with pure vacuum or
static plasma gas. While, if one imagines the very huge plasma gas
corotating in the exterior and takes the limit formally $L\rightarrow \infty 
$, then the wave number of region (3) should be equal to that of region (2), 
$k^{(3)}\rightarrow k_{\pm }^{(2)}$. We thus have that $\varkappa
_{+}\rightarrow 2k_{\pm }^{(2)},\varkappa _{-}\rightarrow 0$ and $\gamma
\rightarrow 1$, which yield the boundary condition $d/dz[\ln \xi _{\pm
}^{(2)}]=ik_{\pm }^{(2)}$.

One can draw some important facts from the boundary condition (\ref{bc at
surf}). Especially when the wave number satisfies the periodic condition $%
\varkappa _{\pm }L=n\pi $ $(n=1,2,\cdots )$, this yields $\gamma
=k^{(3)}/k_{\pm }^{(2)}$ and thereby the boundary condition for \textit{pure
vacuum} is recovered $d/dz[\ln \xi _{\pm }^{(2)}]=ik^{(3)}$. The background
rotation has no influence only on the Alfven waves satisfied with this
condition. Whereas, when the wave number satisfies the another periodic
condition $\varkappa _{\pm }L=(2n-1)\pi /2$ $(n=1,2,\cdots )$, one obtains $%
\gamma =k_{\pm }^{(2)}/k^{(3)}$ and thus the boundary condition works out to
be $d/dz[\ln \xi _{\pm }^{(2)}]=ik^{(3)}(k_{\pm }^{(2)}/k^{(3)})^{2}\gg
ik^{(3)}$. The RHS of this expression turns out that the waves satisfied
with this condition are highly transmitted, as far as $k_{\pm }^{(1)}$ does
not change much over one wavelength, say the validity of WKB approximation $%
\left( 1/k_{\pm }^{(1)}\right) \left| dk_{\pm }^{(1)}/dz\right| \ll k_{\pm
}^{(1)}$ holds. This peculiar result means that the internal shear modes can
be coupled onto the high frequency Alfven modes at the stellar surface,
which drastically enhances their transmission rate at certain wave numbers
or frequencies.

\begin{center}
\bigskip
\end{center}

This work is supported in part by Grants-in-Aid for Scientific Research
(14047215, 16029207, and 16540256) from the Japanese Ministry of Education,
Culture, Sports, Science, and Technology.

\bigskip \bigskip

\end{document}